\documentclass[apj]{emulateapj}

\shorttitle{Three dimensional maps of the SMC}
\shortauthors{Haschke, Grebel \& Duffau}
\slugcomment{submitted: 13 April 2011; accepted: 5 June 2012}

\begin{document}
  \title{Three dimensional maps of the Magellanic Clouds \\using RR~Lyrae Stars and Cepheids \\II. The Small Magellanic Cloud}


\author{Raoul Haschke\altaffilmark{*}
     ,
	 Eva K. Grebel,
	 and
     Sonia Duffau
	  }

\email{haschke@ari.uni-heidelberg.de}

\affil{Astronomisches Rechen-Institut, Zentrum f\"ur Astronomie der Universit\"at Heidelberg,
	   M\"onchhofstra\ss e 12-14, D-69120 Heidelberg, Germany}\

\altaffiltext{*}{Raoul Haschke is a member of the Heidelberg Graduate School for Fundamental Physics (HGSFP) and of the International Max Planck Research School for Astronomy and Cosmic Physics at the University of Heidelberg.}

\begin{abstract}
We use data on variable stars from the Optical Gravitational Lensing Experiment (OGLE~III) survey to determine the three-dimensional structure of the Small Magellanic Cloud (SMC). Deriving individual distances to RR~Lyrae stars and Cepheids we investigate the distribution of these tracers of the old and young population in the SMC.
Photometrically estimated metallicities are used to determine the distances to 1494~RR~Lyrae stars, which have typical ages greater than 9~Gyr. For 2522~Cepheids, with ages of a few tens to a few hundred Myr, distances are calculated using their period-luminosity relation. Individual reddening estimates from the intrinsic color of each star are used to obtain high precision three-dimensional maps.
The distances of RR~Lyrae stars and Cepheids are in very good agreement with each other. The median distance of the RR~Lyrae stars is found to be $61.5 \pm 3.4$~kpc. For the Cepheids a median distance of $63.1 \pm 3.0$~kpc is obtained. Both populations show an extended scale height, with $2.0 \pm 0.4$~kpc for the RR~Lyrae stars and $2.7 \pm 0.3$~kpc for the Cepheids. This confirms the large depth of the SMC suggested by a number of earlier studies. The young population is very differently oriented than the old stars. While we find an inclination angle of $7^\circ \pm 15^\circ$ and a position angle of $83^\circ \pm 21^\circ$ for the RR~Lyrae stars, for the Cepheids an inclination of $74^\circ \pm 9^\circ$ and a position angle of $66^\circ \pm 15^\circ$ is obtained. The RR~Lyrae stars show a fairly homogeneous distribution, while the Cepheids follow roughly the distribution of the bar with their northeastern part being closer to us than the southwestern part of the bar. Interactions between the SMC, LMC, and Milky Way are presumably responsible for the tilted, elongated structure of the young population of the SMC.
\end{abstract}


%

\section{Introduction}
\label{introduction}
%
%
\defcitealias{Haschke11_reddening}{Paper~I}
\defcitealias{Gonidakis09}{G09}
The Small Magellanic Cloud (SMC) is a dwarf irregular satellite of the Milky Way \citet{Bergh99}. It is interacting with its larger companion, the Large Magellanic Cloud (LMC), and with the Milky Way. A number of studies have suggested that the apparently disturbed shape and large depth extent of the SMC were caused by these interactions, although the details and the three-dimensional shape of the SMC remain under debate \citep[e.g.,][]{Bekki09c}.

The distance to the SMC is usually assumed to be 60~kpc or $(m-M)_0 = 18.90$~mag \citep[e.g.,][]{Westerlund97}. But even the usage of a specific distance indicator may still lead to different results in different studies. Using, for instance, eclipsing binaries to determine the mean distance of the SMC \citet{North10} obtained $(m-M)_0 = 19.11 \pm 0.03$~mag, while \citet{Hilditch05} found a mean value of $(m-M)_0 = 18.91 \pm 0.03$~mag in very good agreement with the mean distance quoted above. 

The best option to obtain comparable and trustworthy mean distances and structural parameters from different stellar tracers are large surveys. For these the systematic uncertainties are lower, because they have been observed, reduced and analyzed following a coherent procedure and because they tend to provide excellent number statistics for certain distance indicators. Moreover, such surveys can provide stellar tracers of different and distinct ages, an important prerequisite to resolve the different evolutionary states the galaxy has passed through during its history \citep[see, e.g.,][for a summary of the older distance studies]{Westerlund97}. For the SMC (small) sets of very different stellar tracers, representing young, intermediate-age and old populations, have been investigated for many decades to find a common mean distance, assuming that all these different populations have the same distance from us. The differences between the tracers, but also within the results obtained using a specific tracer are greater than the $1\sigma$-uncertainties of the resulting distances. This may be taken as an indication for a significant depth extent of the SMC, but also shows that it is absolutely necessary to analyze large datasets instead of choosing a small, possibly localized subsample of stars within the SMC.

The mean distance moduli of the young population, traced predominantely by Cepheids, are usually greater than $(m-M)_0 = 18.90$~mag \citep[compare, e.g,][]{Laney86, Groenewegen00, Keller06} but also \citet{Ciechanowska10}. For the old RR~Lyrae stars most of the distance estimates reveal shorter distances than those of the Cepheids. The range of values, however, is large with distance moduli between $(m-M)_0 = 18.78 \pm 0.15$~mag and $(m-M)_0 = 19.20$~mag \citep[see, e.g.,][]{Szewczyk09, Deb10, Kapakos11}. For a compilation of distance moduli we refer to the book of \citet{Westerlund97} and to Table~\ref{distance_table}. For the forthcoming analysis the absolute mean distance is not of crucial importance. It is nontheless an interesting open question.

Apart from differences in the calibration of a given distance indicator, differences in the mean distance may, to some extent, be introduced by a large depth of the SMC as found by, e.g., \citet{Mathewson88}. Using Cepheids they showed that the SMC has a considerable depth of about 20~kpc. Newer estimates using intermediate-age tracers led to lower values of the line-of-sight depth. In \citet{Crowl01} a depth between 6~kpc and 12~kpc, in dependence of reddening, was estimated using cluster distances derived by isochrone fitting. In \citet{Subramanian09} the distribution of red clump (RC) stars was investigated and a depth of less than 5~kpc was found for the SMC, in good agreement with the results of \citet{Subramanian12}. They also investigated the depth extent of the old population by using RR~Lyrae stars and found a mean depth of $4.07 \pm 1.68$~kpc. This is in very good agreement with the $4.13 \pm 0.27$~kpc found for the line-of-sight depth of RR~Lyrae stars by \citet{Kapakos11}. 

\citet{Lah05} conducted a study utilizing different stellar tracers to show that red giant branch (RGB) and asymptotic giant branch (AGB) stars might be further away from us than the main body of the SMC (see their Fig.~4). In \citet{Subramanian12} the inclination angle of the RC stars and the RR~Lyrae stars was found to be $i \simeq 0.5^\circ$. Furthermore they found a position angle of $\theta = 58.3^\circ$ for the RR~Lyrae stars and of $\theta = 55.5^\circ$ for the RC stars, respectively, thus very similar values for old and intermediate-age stars. For the other determinations of the structural parameters of the SMC only young stars have been used. With 63 Cepheids \citet{Caldwell86} found an inclination angle of $i = 70^{\circ} \pm 3^{\circ}$ and a position angle $\theta = 58^{\circ} \pm 10^{\circ}$. With a different sample of 23 Cepheids \citet{Laney86} obtained $i = 45^{\circ} \pm 7^{\circ}$ and a position angle $\theta = 55^{\circ} \pm 17^{\circ}$ in good agreement with \citet{Caldwell86}. A much larger sample of 236 Cepheids was investigated by \citet{Groenewegen00}, who found $i = 68^{\circ} \pm 2^{\circ}$ and a position angle of the line of nodes of $\Theta = 238^{\circ} \pm 7^{\circ}$. 

The location of the center of the SMC is also not very well constrained. The optical center \citep[see, e.g.,][]{Westerlund97} and the center of the K- and M-stars \citep[found by][from now on G09]{Gonidakis09} are basically identical with $\alpha = 0^{\mathrm{h}}51^{\mathrm{m}}$ and $\delta = -73.1^\circ$. From Hubble Space Telescope (HST) measurements of proper motions \citet{Piatek08} found the kinematical center of the SMC to be at $\alpha = 0^{\mathrm{h}}52^{\mathrm{m}}8^{\mathrm{s}}$ and $\delta = -72.5^\circ$, while \citet{Stanimirovic04} found $\alpha = 0^{\mathrm{h}}47^{\mathrm{m}}33^{\mathrm{s}}$ and $\delta = -72^\circ5'26''$ for the highest H~I column density. Throughout this paper we will mostly refer to the optical center and use the result by \citetalias{Gonidakis09}.

In this paper we analyze the data of the Optical Gravitational Lensing Experiment (OGLE~III) survey, presented in Section~\ref{data}. Distances to all RR~Lyrae~\textit{ab} stars and Cepheids present in the OGLE sample are calculated. In Section~\ref{distance} we use the metallicity estimates of \citet{Haschke12_MDF} and the periods obtained by OGLE. Moreover we apply the reddening maps of \citet{Haschke11_reddening} to correct for individual reddening effects. The two-dimensional spatial distribution of the stars is investigated in Section~\ref{Density_of_OGLEIII} and the three-dimensional maps are presented in Section~\ref{3D}. These three-dimensional maps are analyzed and the structural parameters of the young and old population are determined in Section~\ref{3D_structure}. The results are discussed and summarized in Section~\ref{Conclusions}.

%

\section{Data}
\label{data}

In 2001 the OGLE experiment started its third phase of monitoring the Magellanic Clouds (OGLE~III). This phase ended in 2009. OGLE~III used a camera of eight CCDs with $2048 \times 4096$ pixels each and a field of view of to $35' \times 35'$. Altogether 14 square degrees, covering the bar and the wing of the SMC, were monitored. Photometric data in the $V$ and $I$~band were accumulated for 6.2~million stars \citep{Udalski08b}. Apart from the full photometric catalog, the OGLE collaboration provides specialized catalogs with information about certain types of stars, such as Cepheids or $\delta$~Scuti stars, or astrometric properties of stars\footnote{The catalogs are available at \url{http://ogle.astrouw.edu.pl/}}.

In \citet{Soszynski10a} the data for 2626 classical Cepheids in the SMC are presented, while \citet{Soszynski10b} published data for 1933 RR~Lyrae stars of type \textit{ab}. All of these stars are pulsating in the fundamental mode and cover the entire OGLE~III field of the SMC. The lightcurves were analyzed and periods and mean magnitudes in V- and I-band were published. The OGLE collaboration also carried out a Fourier decomposition \citep{Simon93} of the very well sampled I-band lightcurves. The Fourier parameters $R_{21}$ and $R_{31}$, which correspond to the skewness of the lightcurve, as well as $\phi_{21}$ and $\phi_{31}$, which represent the acuteness \citep[see][for more details]{Stellingwerf87a}, of each Cepheid and RR~Lyrae are published and available from the OGLE website.

%

\section{Distance measurements}
\label{distance}

\begin{table*}
\caption{Compilation of distance estimates for the SMC from different tracers using recent literature values and this work.} 
\label{distance_table} 
\centering 
\begin{tabular}{r c l} \hline \hline
Type of indicator & mean distance & Reference \\ 
& $(m-M)_0 \pm \sigma$ & \\ \hline 
Cepheids & $19.11 \pm 0.11$ & \citet{Bono01} \\
Cepheids & $18.85 \pm 0.14$ & \citet{Ciechanowska10} \\
Cepheids & $19.11 \pm 0.11$ & \citet{Groenewegen00b} \\
Cepheids & $18.93 \pm 0.024$ & \citet{Keller06} \\
Cepheids & $19.17 \pm 0.12$ & this work - area-averaged reddening \\ 
Cepheids & $19.00 \pm 0.10$ & this work - individual reddening \\ 
CMD fitting & $18.88 \pm 0.08$ & \citet{Dolphin01} \\
Eclipsing binaries & $18.89 \pm 0.14$ & \citet{Harries03} \\
Eclipsing binaries & $18.91 \pm 0.13$ & \citet{Hilditch05} \\
Eclipsing binaries & $19.11 \pm 0.03$ & \citet{North10} \\
RGB tip & $18.99 \pm 0.11$ & \citet{Cioni00b} \\
RR~Lyrae & $18.86 \pm 0.01$ & \citet{Deb10} \\
RR~Lyrae~\textit{ab} & $18.90 \pm 0.18$ & \citet{Kapakos11} \\
RR~Lyrae~\textit{c} & $18.97 \pm 0.14$ & \citet{Kapakos11} \\
RR~Lyrae & $18.97 \pm 0.15$ & \citet{Szewczyk09}\\
RR~Lyrae & $19.13 \pm 0.13$ & this work - area-averaged reddening \\ 
RR~Lyrae & $18.94 \pm 0.11$ & this work - individual reddening \\ \hline 
\end{tabular}
\end{table*}

In this Section we discuss our distances to RR~Lyrae stars as tracers of the old population \citep[$\geq 9$~Gyr, e.g.,][]{Sarajedini06} and to Cepheids for the young population \citep[$ \sim 30-300$~Myr,][]{Grebel98, Luck03} of the SMC. Distance estimates are calculated individually star by star.

\subsection{RR~Lyrae}

In \citet{Haschke12_MDF} the photometric metallicities of 1864 RR~Lyrae~\textit{ab} stars from the OGLE~III sample of the SMC were calculated. The absolute luminosity of RR~Lyrae stars depends on metallicity only. We use these estimates on the metallicity scale of \citet{Zinn84} to calculate the absolute $V$ band magnitude $M_V$ of the RR~Lyrae stars by using the relation introduced in \citet{Benedict11}

\begin{equation}
M_V = (0.45 \pm 0.05) + (0.217 \pm 0.047)~(\textrm{[Fe/H]} + 1.5)
\label{absolute_magnitude}
\end{equation}

\citet{Benedict11} found that their {\it HST} data are fitted best by an equation having the same slope as the relation found by \citet{Clementini03} for LMC RR~Lyrae stars, but with a zeropoint that is $\Delta M_V = 0.07$~mag brighter.

Furthermore, we also tried out quadratic equations by \citet{Catelan04, Sandage06} and \citet{Bono07}. The resulting mean absolute magnitudes of the RR~Lyrae stars are quite different from each other. We find median differences of $\Delta M_V = 0.10$~mag for \citet[their equation~7]{Sandage06}, $\Delta M_V = 0.17$~mag for \citet[their equation~8]{Catelan04} and $\Delta M_V = 0.20$~mag for \citet[their equation~10]{Bono07}. All of these absolute magnitudes are fainter than those calculated with the  relation by \citet{Benedict11}. This reflects the comparatively large uncertainty of the magnitude zeropoint of the RR~Lyrae stars. We take these systematics for our error analysis into account, and adopt the relation of \citet{Benedict11}.

The absolute magnitude of each RR~Lyrae star together with the observed mean magnitude from the OGLE collaboration yields a distance modulus once a reddening correction is applied. In Section~\ref{reddening} we apply two different approaches to correct for the reddening and to obtain the three-dimensional structure of the SMC.

\subsection{Cepheids}

The correlation between absolute magnitude and period is well known for Cepheids. Using Cepheids from the SMC \citet{Sandage09} obtained relations for the $B$, $V$, and $I$ band, which only depend on the period $P$ of the Cepheid investigated: 

\begin{eqnarray}
M_V = -(2.588 \pm 0.045)\log P - (1.400 \pm 0.035) \label{Cepheid_M_V} \\
M_I = -(2.862 \pm 0.035)\log P - (1.847 \pm 0.027) \label{Cepheid_M_I}
\end{eqnarray}

There might be a break in the relation for the SMC at $P = 10$~days as found for the LMC \citep{Sandage04a}, but \citet{Sandage09} suggest that this break is not significant. 

The distance moduli are calculated using the mean observed magnitude of each star, provided by the OGLE collaboration, and the absolute magnitude from these relations. The correction for reddening effects is described in the next section.

\subsection{Reddening correction}
\label{reddening}

\citet[hereafter Paper~I]{Haschke11_reddening} used two different approaches to correct for the reddening of the LMC. We use the same method here, which will be shortly outlined. Further details are described in \citetalias{Haschke11_reddening}.

\subsubsection{Area-averaged reddening corrections:\\The red clump method}
RC stars are located at a certain color and luminosity range of the color-magnitude diagram (CMD). That position depends on the distance, the metallicity, and the reddening of these stars. \citet{Girardi01} predicted the theoretical color of the RC for distinct metallicities using models. Assuming a metallicity of $z = 0.025$ \citep{Cole98, Glatt08b} for the SMC RC population the difference of the observed to the theoretically predicted color provides the reddening \citep{Wozniak96}.

In \citet{Haschke11_reddening}, we estimated reddening values $E(V-I)$ for 681 subfields in the OGLE~III field of the SMC. These were converted using the relation by \citet{Schlegel98} 

\begin{eqnarray}
A_V = 3.24(E(V-I)/1.4) \label{reddening_V} \\
A_I = 1.96(E(V-I)/1.4) \label{reddening_I}
\end{eqnarray}

to correct the apparent magnitudes of the stars with the mean extinction in the corresponding field.

\subsubsection{Individual reddening correction:\\Intrinsic colors of variable stars}
\label{reddening_color}

Individual reddening values of RR~Lyrae stars and Cepheids were calculated by subtracting the intrinsic color from the observed color of each star. The observed color $(v-i)$ was computed from the mean magnitudes of the OGLE data. For the intrinsic color $(V-I)_0$, we calculated the absolute magnitudes in $V$ and $I$ of each RR~Lyrae star using the relations of \citet{Catelan04}. For the absolute $V$ and $I$ magnitudes of each Cepheid Equation~\ref{Cepheid_M_V} and Equation~\ref{Cepheid_M_I} were used.

The reddening estimates were transformed to individual extinction values using Equation~\ref{reddening_V} and Equation~\ref{reddening_I} and the reddening-free distances were calculated. Individual extinction corrections calculated for each target star separately have the advantage of not being subject to unaccounted differential reddening nor population effects \citep[see also][]{Zaritsky02}. For a detailed description of the method we refer the interested reader to \citetalias{Haschke11_reddening}.

%

\section{Star densities in the OGLE~III field}
\label{Density_of_OGLEIII}

\begin{figure}
\centering 
 \includegraphics[width=0.47\textwidth]{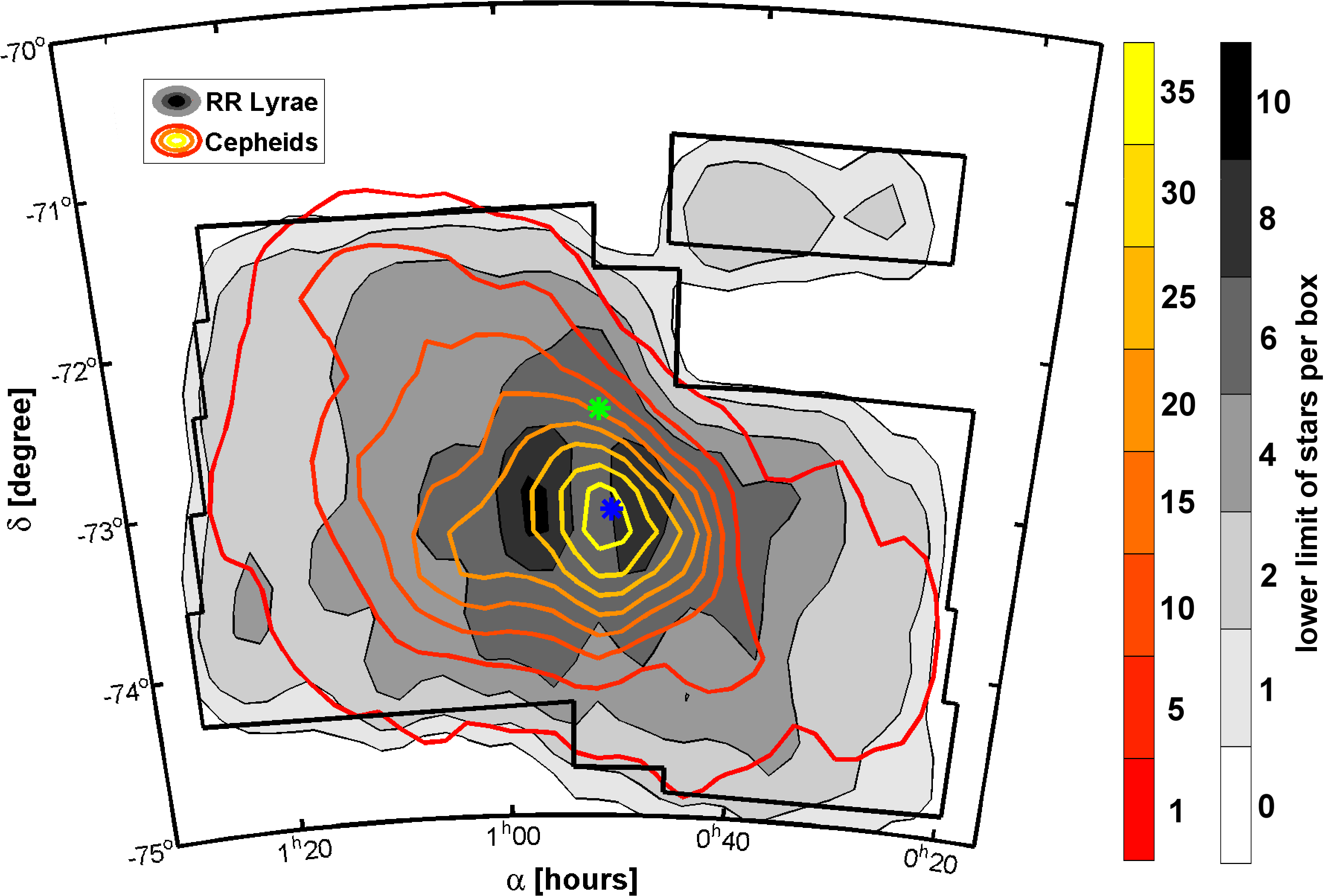}
 \caption{Densities of RR~Lyrae stars (filled grey contours) and Cepheids (colored contours) are shown as a function of right ascension, $\alpha$ (J2000), and declination, $\delta$ (J2000). While the RR~Lyrae stars show a bimodal distribution, the density of Cepheids decreases with distance from the center found by \citetalias{Gonidakis09} (marked with a blue asterisk). The green asterisk represents the kinematic center found by \citet{Piatek08} using {\em HST} proper motions. The box sizes of the evaluated fields are listed in Table~\ref{table_bins_RC_contour}.} 
 \label{RRL_Cep_RADEC_colorext}
\end{figure}

The central parts of the SMC, including the bar and a very small part of the wing of the SMC, are covered by the field of OGLE~III. Overall 14 square degrees are covered and Figure~\ref{RRL_Cep_RADEC_colorext} shows the density distribution of RR~Lyrae stars and Cepheids in the observed field. The stars are counted in boxes of $0.5^{\circ} \times 0.25^{\circ}$ in $\alpha$ and $\delta$, respectively, (see Table~\ref{table_bins_RC_contour}) and the resulting distributions are smoothed using a Gaussian kernel. 

For the RR~Lyrae stars the distribution is very smooth and increases steadily from the outskirts towards the center found by \citetalias{Gonidakis09} from K- and M-stars. However, the highest density of RR~Lyrae stars is not at the center, but the peak of the distribution is off-centered and nearly bimodal, as shown in Figure~\ref{RRL_Cep_RADEC_colorext} \citep[see also Figure~7 in][]{Soszynski10b}. We may expect that a considerable number of RR~Lyrae stars is located outside of the field of view of OGLE~III, especially towards the south and northwestern direction. In the northwest OGLE~III targeted an extended rectangular region where also a number of SMC RR~Lyrae stars were detected (separate rectangular area in the northwest in Figure~\ref{RRL_Cep_RADEC_colorext}), as well as the region of the Galactic globular cluster 47~Tuc. For distance estimates with the RC reddening correction the stars in the field of the cluster 47~Tuc are excluded, while they are taken into account for the individual reddening-corrected distance estimates.

The distribution of the Cepheids is very different from that of the old population traced by the RR~Lyrae stars. Very close to the center of \citetalias{Gonidakis09} the density of Cepheids is highest (Figure~\ref{RRL_Cep_RADEC_colorext}). Then it drops with increasing distance from the center. First the isodensity lines are nearly circular, but with increasing distance from the center the isodensity contours become more elongated towards the west and northeastern directions. Interestingly no Cepheids are found in the most northern fields, as well as at the position of 47~Tuc in agreement with \citet{Graham75} (47 Tuc is not indicated in Figure~\ref{RRL_Cep_RADEC_colorext}). Recent star formation has therefore not taken place in these outer regions of the SMC OGLE~III field, but is more strongly concentrated in the area of the bar of the SMC. Note that most of the SMC wing \citep{Shapley40} is not covered by OGLE~III. The wing stands out prominently in H$\alpha$ images and may also contain a larger number of Cepheids. Differences between the distribution of the young and the old populations in the SMC using other stellar tracers can also be seen in \citet{Zaritsky00}.

\begin{table}
\caption{Binsizes of the fields evaluated to obtain the densities of Cepheids and RR~Lyrae stars in the SMC.} 
\label{table_bins_RC_contour}      
\centering             
\begin{tabular}{c c c}    
\hline\hline         
 & RR~Lyrae & Cepheids \\  
\hline            
$\alpha$ contour bin [degree] & 0.5 & 0.5 \\ 
$\delta$ contour bin [degree] & 0.25 & 0.25 \\ 
distance contour bin [kpc] & 0.5 & 0.5 \\ 
$\alpha$ isodensity bin [degree] & 2 & 2 \\ 
$\delta$ isodensity bin [degree] & 1 & 1 \\ 
distance isodensity bin [kpc] & 2 & 2 \\ 
\hline                  
\end{tabular}
\end{table}

%

\section{Three dimensional maps}
\label{3D}

For each RR~Lyrae star and Cepheid in our sample distances are calculated using the relations described in Section~\ref{distance}. Either the averaged reddening from the RC stars or the individual reddening method is applied. This results in two independent sets of distance maps for each population.

\subsection{Maps corrected with area-averaged reddening}
\label{corrected_RC}

In this subsection the area-averaged reddening values obtained from the RC stars are used. All RR~Lyrae stars and Cepheids located in one subfield defined by the RC reddening method are extinction-corrected using the same RC reddening value (for details see \citetalias{Haschke11_reddening}). 

\begin{figure}
\centering 
 \includegraphics[width=0.47\textwidth]{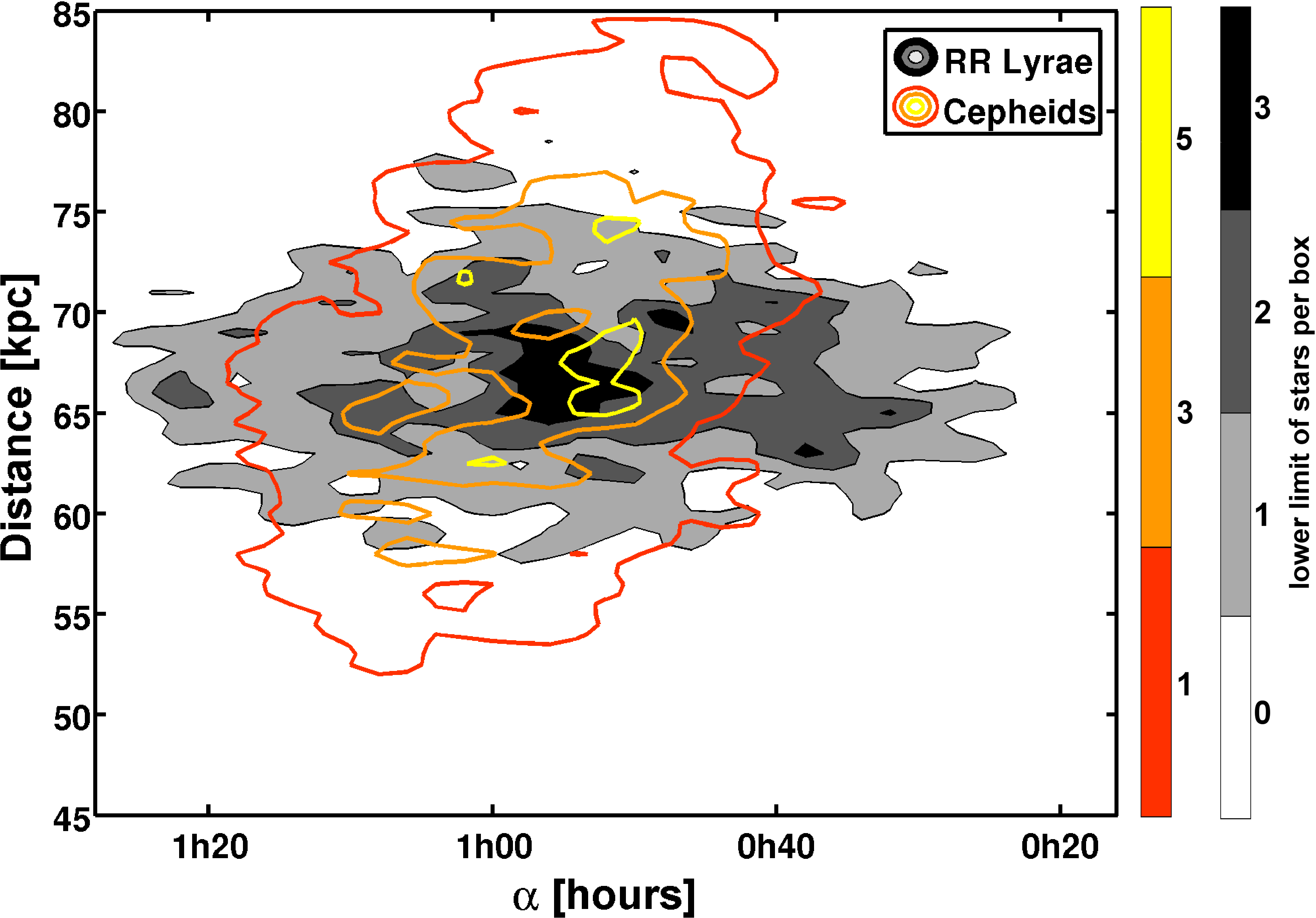}
\vspace{\floatsep} 
 \includegraphics[width=0.47\textwidth]{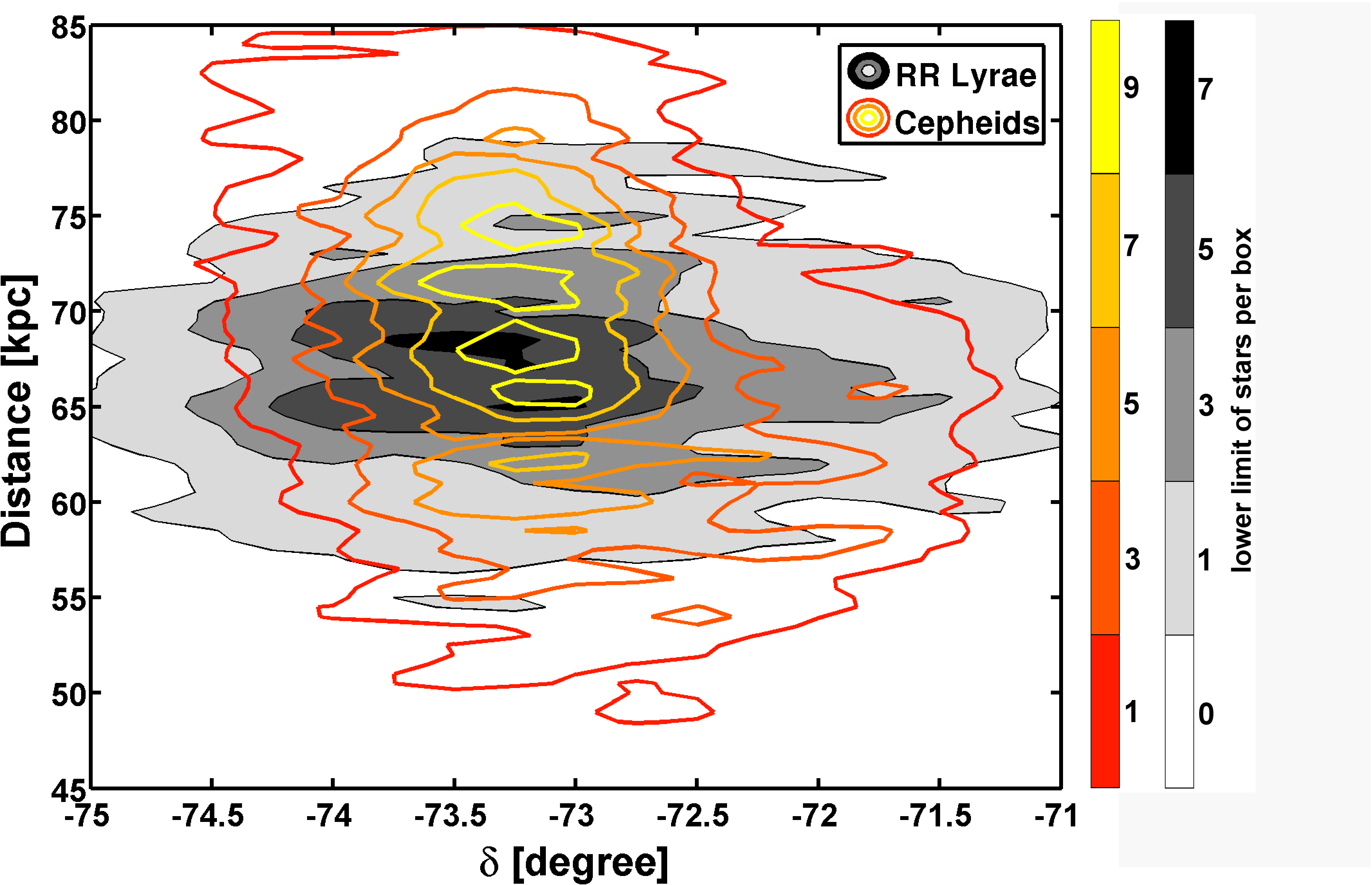}
 \caption{Stellar densities of RR~Lyrae stars (filled grey contours) and Cepheids (colored contours) shown as a function of distance and right ascension $\alpha$ in the upper panel, and as a function of distance and declination $\delta$ in the lower panel. Area-averaged reddening values are used to correct all distance estimates. The main concentration of both populations is located nearly at the same position. But the inclination angles of the young and old population are very different from each other. In Table~\ref{table_bins_RC_contour} the sizes of the evaluated boxes are listed.} 
 \label{RRL_Cep_dist_RCext}
\end{figure}

In Figure~\ref{RRL_Cep_dist_RCext} we plot the spatial position of the RR~Lyrae stars (grey contours) and Cepheids (colored contours) in $\alpha$ and $\delta$, respectively, versus distance. This corresponds to a change of the viewing direction of the observer to a northern position above the SMC in the upper panel, and to the eastern side in the lower panel. Table~\ref{table_bins_RC_contour} lists the box sizes within which the density of stars is evaluated. In order to smooth the density contour plot a Gaussian kernel with a width of $3 \times 3$~bins is used. The variances on very small scales are reduced by this procedure. 

The location of the central concentrations of the two populations coincide roughly in Figure~\ref{RRL_Cep_dist_RCext}. The RR~Lyrae stars have a median distance of $D_{\mathrm{RRL/median}} = 66.8 \pm 4.3$~kpc ($(m-M)_0 = 19.13 \pm 0.13$~mag). The median distance of the Cepheids is essentially the same with a distance of $D_{\mathrm{Cep/median}} = 68.1 \pm 4.1$~kpc ($(m-M)_0 = 19.17 \pm 0.12$~mag). For the uncertainties we take the mean magnitude error of 0.07~mag, stated by the OGLE collaboration, the error of the metallicity of 0.23~dex, as determined by \citet{Haschke12_MDF} and the mean extinction error of 0.08~mag, found by \citetalias{Haschke11_reddening}, and compute the uncertainty by using error propagation. The error of the period as stated by OGLE is too small to be a contributing factor. 

While our mean distances of the old and young population of the SMC are in very good agreement, this could in principle be attributable to the initially chosen zeropoints. Nonetheless we do not think that it is purely furtuitous. In contrast to many earlier studies our tracers were taken from a homogeneous database, taken with the same instrument and filters and reduced with the same procedure. Additionally, we use state-of-the-art relations to calculate distances. We realize that recent determinations of zeropoints and slopes in the literature no longer show major differences, hence we think that an intrinsic zeropoint shift between the different tracers is not likely. We believe that the lack of a measurable offset between the two populations does imply that their mean distances do coincide.

\subsection{Maps corrected with individual reddening}

A more precise reddening correction can be applied by using individual reddening estimates for each Cepheid or RR~Lyrae star, instead of using an area-averaged reddening value for a large number of RC stars. We use the reddening estimates derived in \citetalias{Haschke11_reddening} to correct for the individual intrinsic color differences. The color from the absolute magnitudes in the $V$ and $I$ band are compared with the color from the observed $v$ and $i$ apparent magnitudes. The difference is assumed to be the individual reddening of this star, as described in Section~\ref{reddening}. 

\begin{figure}
\centering 
 \includegraphics[width=0.46\textwidth]{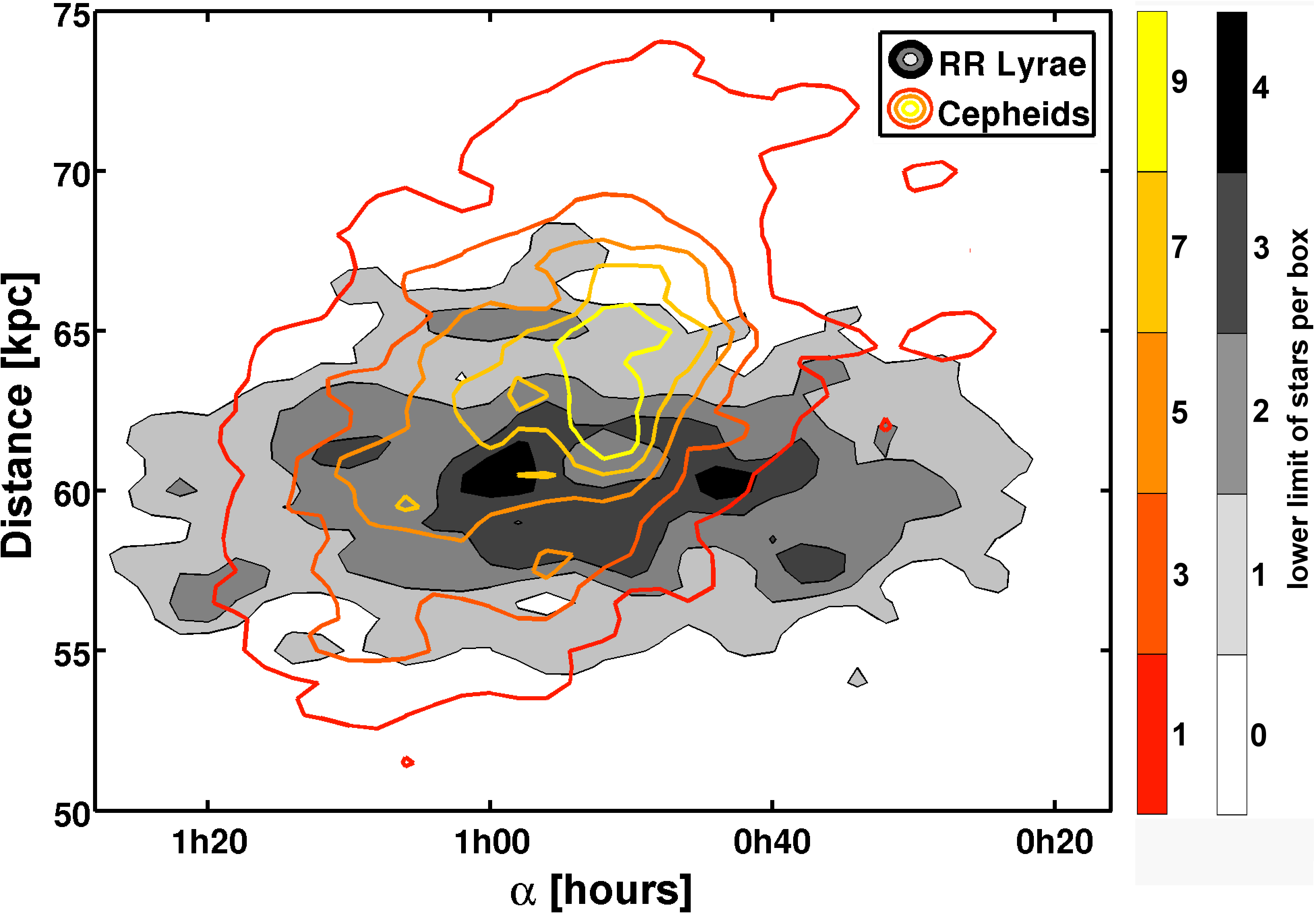}
\vspace{\floatsep}
 \includegraphics[width=0.46\textwidth]{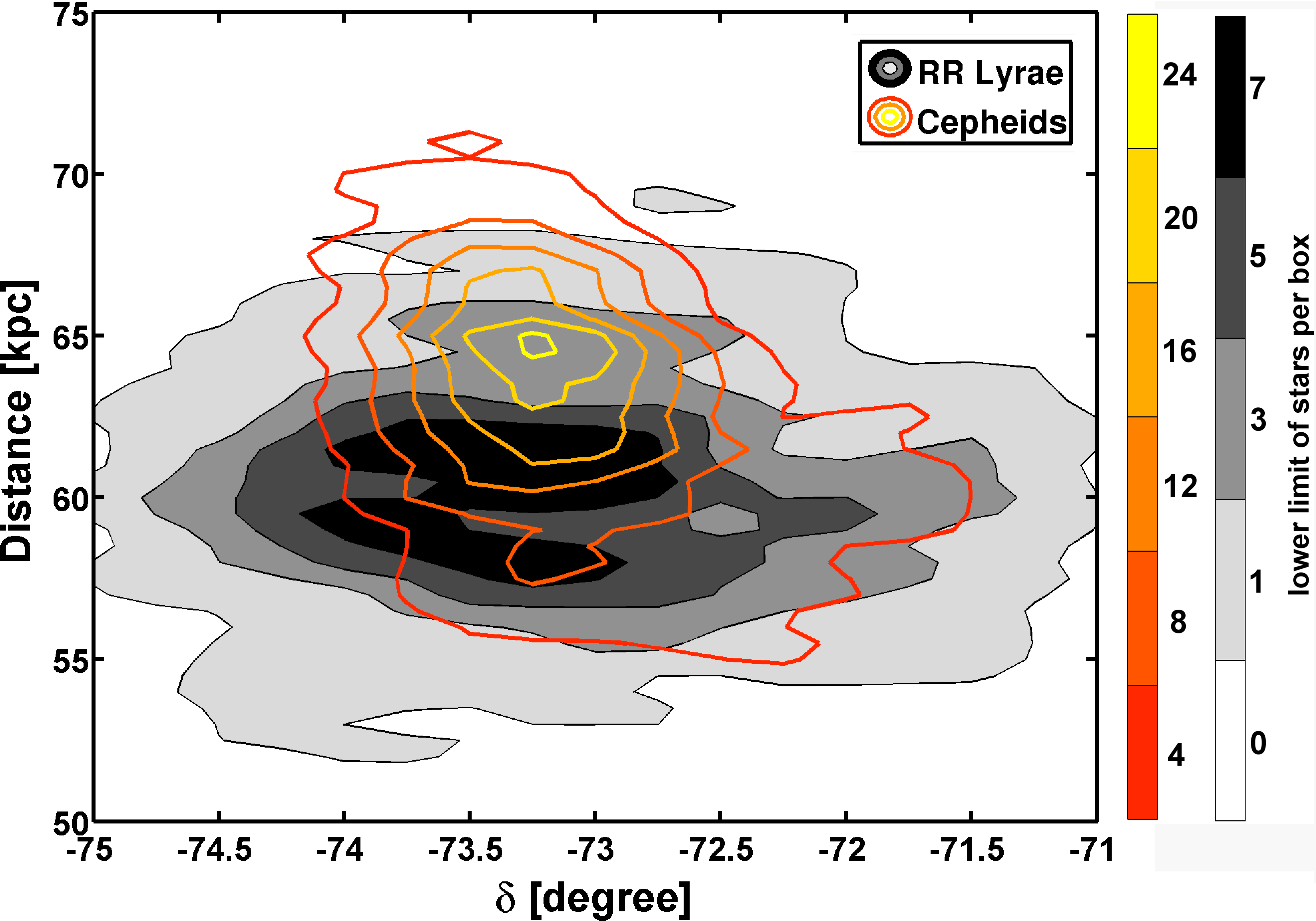}
\vspace{\floatsep}
 \includegraphics[width=0.41\textwidth]{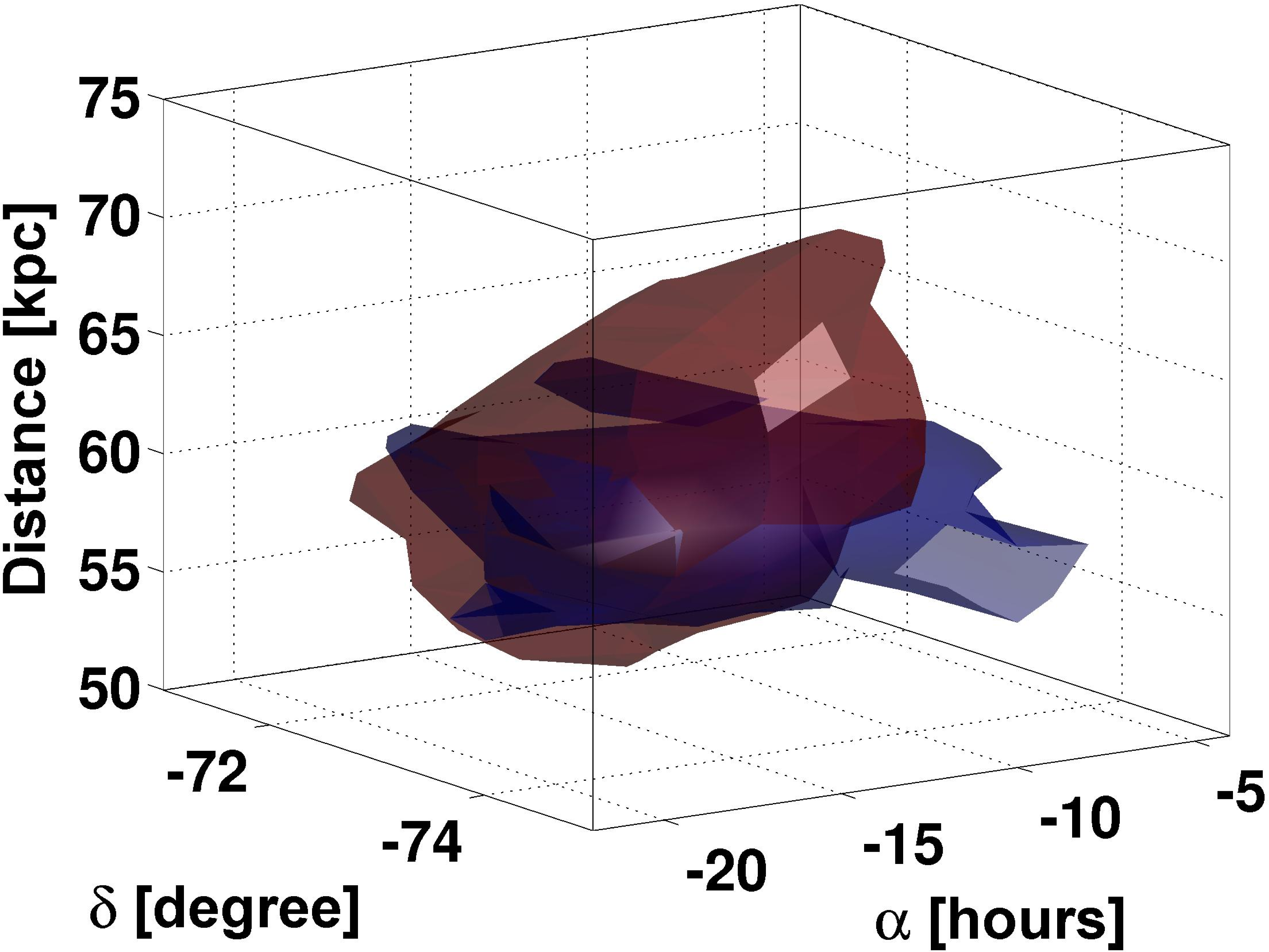}
 \caption{Stellar densities of RR~Lyrae stars (filled grey contours) and Cepheids (colored contours) as a function of distance and right ascension $\alpha$ in the upper panel and as a function of distance and declination $\delta$ in the middle one. The lower panel shows a three-dimensional representation of an isodensity contour of the RR~Lyrae stars (blue) and Cepheids (red) as a function of right ascension $\alpha$, declination $\delta$, and distance. All distances are extinction-corrected using the individual color reddening values. The distributions of the RR~Lyrae stars and the Cepheids have a very different orientation in the SMC. While the RR~Lyrae form a flattened disk-like structure and are not inclined, the Cepheids show a large inclination angle. Furthermore, the RR~Lyrae show a lower density pattern in the center of the SMC surrounded by a higher density ring (upper panel), which is not visible for the Cepheids. The sizes of the boxes used to evaluate the density are listed in Table~\ref{table_bins_RC_contour}. [The lower panel is also available as an mpeg animation \textit{video1.mpeg} in the electronic version of this article published in the Astronomical Journal. The video shows a 360$^\circ$ rotation of the isodensity contours.]} 
 \label{RRL_Cep_dist_colorext} 
\end{figure}

Figure~\ref{RRL_Cep_dist_colorext} reveals shorter distances for the SMC than when using the RC reddening values in Figure~\ref{RRL_Cep_dist_RCext}. The median distance of the RR~Lyrae stars is found to be $D_{\mathrm{RRL/median}} = 61.5 \pm 3.4$~kpc ($(m-M)_0 = 18.94 \pm 0.11$~mag), a bit closer than using the RC reddening method. For the Cepheids we also find a closer distance using the individual reddening values, $D_{\mathrm{Cep/median}} = 63.1 \pm 3.0$~kpc ($(m-M)_0 = 19.00 \pm 0.10$~mag).

As before in Section~\ref{corrected_RC} we calculate the uncertainties of the distances using error propagation. We take into account the intrinsic magnitude error, the uncertainty of the metallicity estimate plus the uncertainty of the reddening. The period is measured with such high accuracy by OGLE~III that it does not influence the uncertainty determination. 

In \citetalias{Haschke11_reddening}, we point out that in regions with substantial amounts of dust and gas the reddening can fluctuate significantly with depth and position. Such local differential reddening cannot be resolved by the RC reddening method. Furthermore, as pointed out by \citet{Barmby11}, Cepheids suffer from considerable mass loss, leading to circumstellar dust around the star. This leads to additional differential reddening, which is not accounted for by the RC maps. Details on the differences of population- and temperature-dependent reddening are discussed in \citet{Grebel95, Zaritsky99} and \citet{Zaritsky02}. Details on the differences of population- and temperature-dependent reddening are discussed in \citet{Grebel95, Zaritsky99} and \citet{Zaritsky02}. A more accurate distance estimate can be obtained by using the individual reddening method.

The mean distances of the RR~Lyrae and Cepheid population are in good agreement within their uncertainties when using individual dereddening. As found when using the RC reddening the SMC has a considerable depth, even though it is reduced, as expected, when using the individual reddening. 

The internal structure of the SMC changes as well depending on the kind of reddening correction. In the upper panel of Figure~\ref{RRL_Cep_dist_colorext} we obtain a lower density for the RR~Lyrae stars in the center of the SMC. Moving away from the center of the RR~Lyrae distribution, a ring-like structure of higher density surrounds the center. Further outwards the density drops steadily. We check whether this pattern is an artifact introduced by the smoothing with the Gaussian kernel, but the unsmoothed figure contains the same pattern as shown in this representation. By plotting single stars instead of contours we find the same effect and thus consider it to be real. This bimodal distribution is also seen in Figure~7 of \citet{Soszynski10b}. For the Cepheids we find a more centrally concentrated distribution when using the individual stellar reddening. The region of the highest density coincides in part with the low density center of the RR~Lyrae (see Figure~\ref{RRL_Cep_dist_colorext}) and with the center of \citetalias{Gonidakis09}. With increasing distance from the center, the density of the Cepheids drops and the density contours become increasingly elongated. Towards the northeast of the OGLE field the density contours are closer to us and are elongated in the direction towards the Magellanic Bridge.

%

\section{Three-dimensional structure}
\label{3D_structure}

\subsection{The SMC in slices}

The three-dimensional data for the two populations traced by the Cepheids and RR~Lyrae stars allow us to gain a better insight into the internal structural properties of the SMC. In the Figures~\ref{RRL_dist_45_75} and \ref{Cep_dist_45_75} we slice the SMC into three bins of 10~kpc depth each. All stars are color-coded based on their distance and plotted with their spatial coordinates. The individual distance uncertainties of each star are about 8\%, therefore well below the binsize.

\subsubsection{RR~Lyrae stars}

\begin{figure*}
\centering 
 \includegraphics[height=0.67\textheight,width=1\textwidth]{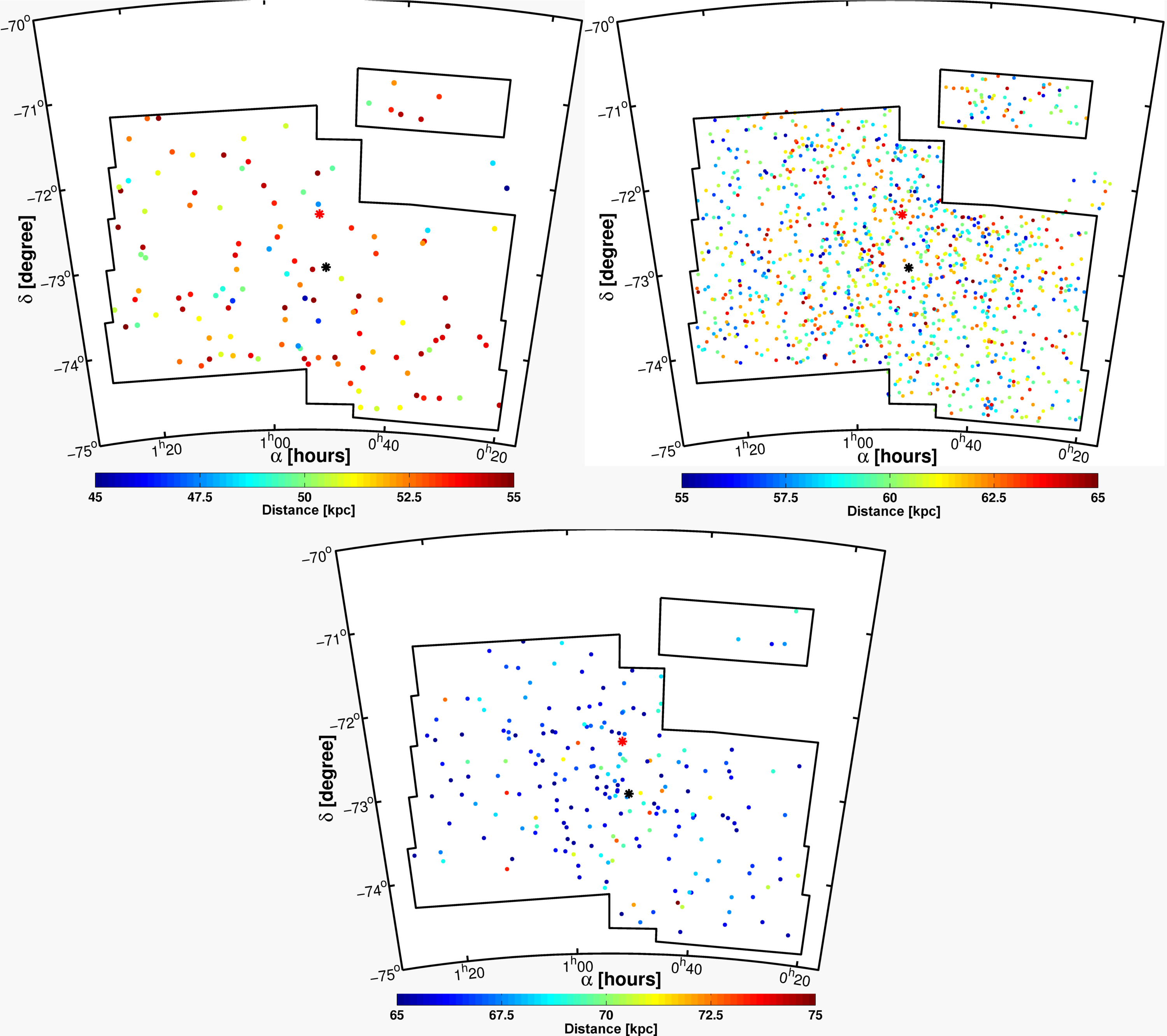}
 \caption{Color-coded distance map for individual RR~Lyrae stars in three SMC distance bins. The distances are sliced into three distance bins of 10~kpc each, covering 45 to 75~kpc. The closest bin (upper left panel) has hardly any stars and no patterns are visible. For the other two bins we cannot see much substructure. The stars seem to be fairly evenly distributed. The black asterisk represents the center found by \citetalias{Gonidakis09}. The upper left panel contains 119~stars, the upper right one 1198 and the lower one 203~RR~Lyrae stars.}
 \label{RRL_dist_45_75} 
\end{figure*}

The projected distribution of the old population of RR~Lyrae stars is shown in Figure~\ref{RRL_dist_45_75} for three different distance bins. In the upper left panel of Figure~\ref{RRL_dist_45_75}, which represents RR~Lyrae stars that are closer than 55~kpc, we find only very few stars of this old generation, most of which are located in the eastern part of the OGLE~III field. Only 1.5\% of all RR~Lyrae stars are present in this distance bin. These stars are randomly distributed and no pattern or structure is visible. 

The intermediate distance bin from 55~kpc to 65~kpc (upper right panel) contains 52.6\% of the whole RR~Lyrae sample. This distance bin is dominated by stars with distances between 60~kpc and 65~kpc. We find that no particular substructure is visible other than a slightly higher density in the central region. The stars are distributed fairly homogeneously over the whole body of the SMC measured by OGLE~III. 

The farthest bin of 65~kpc to 75~kpc contains 45.9\% of the RR~Lyrae stars and is dominated by stars at distances between 65~kpc and 70~kpc. Overall no heterogenous structural patterns are visible. The density distribution of this panel is similar to the intermediate distance bin. 

We test if the density distribution follows a random distribution using the $\mathcal{Q}$-parameter \citep{Cartwright04}. Using the whole sample of RR~Lyrae stars, we find a value of $\mathcal{Q} = 0.73$, which corresponds to a value of a perfect random distribution \citep{Schmeja08}. Dividing the sample in distance bins of 100 RR~Lyrae stars each, values of $\mathcal{Q} = 0.72 - 0.83$ are obtained. Four out of 15 bins have values larger than $\mathcal{Q} > 0.80$, which corresponds to a slight indication for a centrally concentrated distribution, while a mean of $\mathcal{Q} = 0.77 \pm 0.03$ is found. We thus conclude that the RR~Lyrae stars are overall homogeneously distributed over the OGLE~III field of the SMC, showing a slightly ellipsoidal or spheroidal distribution as also found by \citet{Subramanian12}.


\subsubsection{Cepheids}

\begin{figure*}
\centering 
 \includegraphics[height=0.67\textheight,width=1\textwidth]{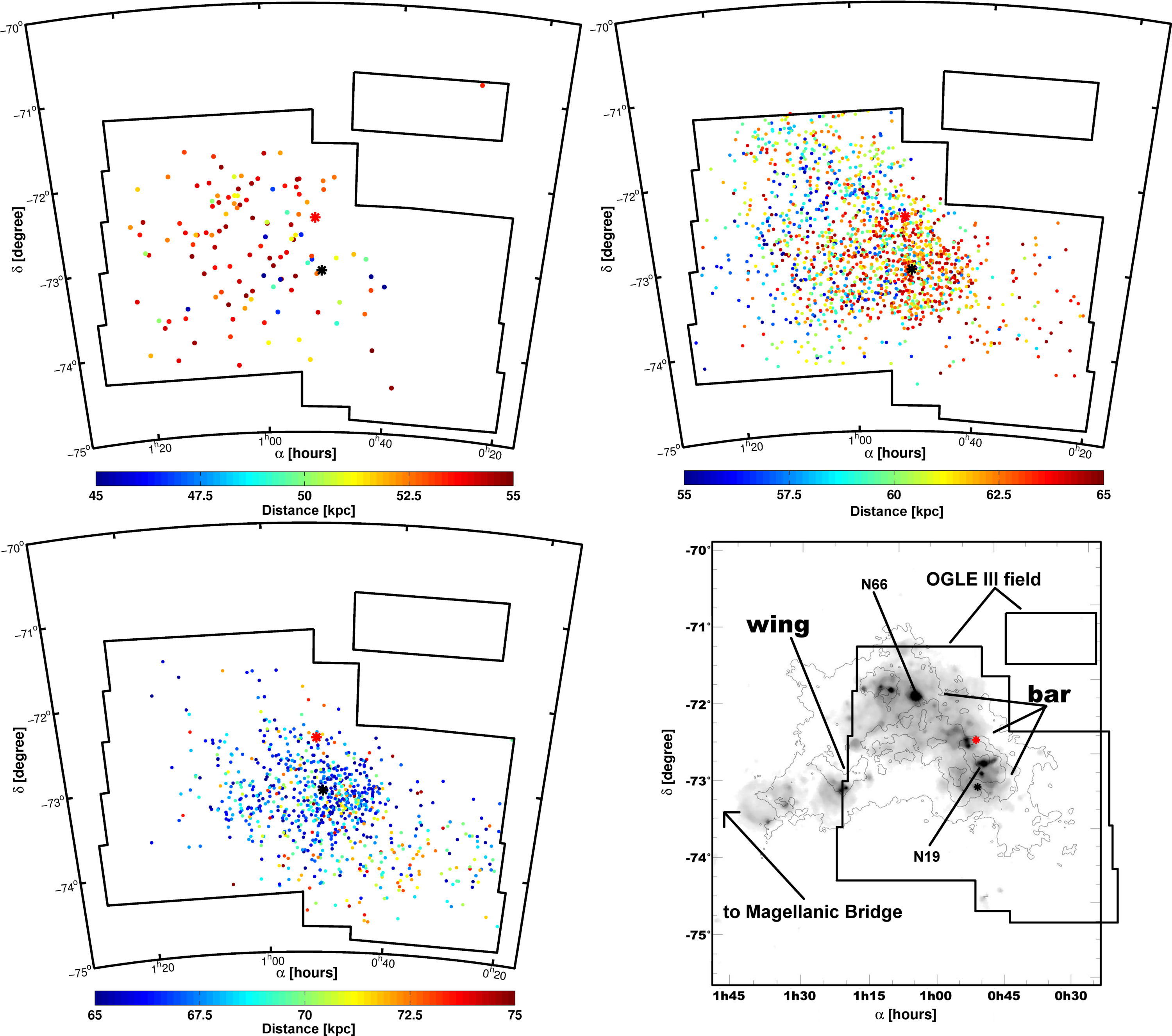}
 \caption{Colour-coded distance map of the Cepheids in the SMC. The stars are shown in three bins of 10~kpc depth each. The upper left panel, representing the closest bin, shows that here most of the Cepheids are concentrated in the eastern regions of the SMC. The northern/northeastern parts of the bar are in general closer than the parts in the southwest. The lower right panel shows the distribution of H$\alpha$ emission \citep{Gaustad01} in greyscale and an H~I contour map \citep{Stanimirovic99}. The combined H$\alpha$-H~I background image in this panel is used with kind permission of Lorimer et al. (NRAO/AUI/NSF). The black asterisk indicates the center of the SMC as found by \citetalias{Gonidakis09}, while the red asterisk represents the center found by \citet{Piatek08}. The main concentration of Cepheids coincides with the center of \citetalias{Gonidakis09}. The most recent star formation has taken place along of the bar, as shown by the H~$\alpha$ contours in the lower right panel. The outer regions of the SMC covered by OGLE~III are devoid of young stars. The panels are populated with 122, 1585 and 815~stars, respectively, when moving from the closest to the farthest bin.} 
 \label{Cep_dist_45_75} 
\end{figure*}
Figure~\ref{Cep_dist_45_75} shows distance bins covering the distance range from 45~kpc to 75~kpc of 10~kpc depth each. The distribution of the Cepheids in these three slices shows a very different picture than for the old population. The distance bin of 45~kpc to 55~kpc contains 122 stars or $5\%$ of the entire sample of the Cepheids. Most of these stars are concentrated in the eastern parts of this closest distance bin. 

The central distance bin of the SMC with a distance range from 55~kpc to 65~kpc (upper right panel) contains the majority, nearly $63\%$, of Cepheids. The panel is dominated by a distance gradient stretching from the northeast to the southwest. The northeastern parts are closest to us and coincident in projection with the region of the SMC bar that contains the luminous N66 H~{\sc ii} region\footnote{The \textquotedblleft N\textquotedblright \ designation of regions luminous in H$\alpha$ follow the numbering scheme introduced in the catalog of \citet{Henize56}.}. In the most eastern part of the bar around N80 we hardly find any Cepheids. The highest density of the Cepheids coincides with the center found by \citetalias{Gonidakis09} very well. This area coincides with a number of H~{\sc ii} regions in the bar, e.g., N17, N19, N20, N22, and N26. The most southernwestern and the most northwestern parts of the OGLE~III region do not contain any Cepheids. 

The farthest bin, with stars at a distance of 65~kpc to 75~kpc, contains a concentration of stars in the projected central region of the OGLE~III field. The eastern parts are nearly devoid of stars in this distance bin and towards the southwest the most distant Cepheids of the SMC are present. Most of the stars concentrated in the central parts of this panel have a similar distance, which is between 65~kpc and 70~kpc. In the northwest of this concentration the density of Cepheids drops drastically and in the most northwestern fields no stars of this young population are present. Recent star formation as traced by Cepheids only took place at locations close to the bar of the SMC ($\alpha \sim 12^{\circ}$ to $16^{\circ}$ and $\delta \sim -73.5^\circ$ to $-72^\circ$).

The lower right panel in Figure~\ref{Cep_dist_45_75} shows the H~I contours \citep[observed by][]{Stanimirovic99} overlayed with observations of H$\alpha$ by \citet{Gaustad01} in greyscale. The combined H$\alpha$-H~I image was taken from Lorimer et al. (NRAO/AUI/NSF)\footnote{http://www.nrao.edu/pr/2007/brightburst}. The Cepheids trace the location of the bar very well. The highest density of visible light agrees with the highest density of Cepheids visible in the upper right and lower left panel. As mentioned in the previous section, the northeastern region of the OGLE~III field contains the Cepheids with the closest distance to us. This is in good agreement with the inclined gaseous component of the SMC found by \citet{Stanimirovic04}. The number density of Cepheids decreases from the bar in the direction towards the wing. The wing itself is missing from the OGLE~III field, which ends just west of the H~{\sc ii} region N84.

\subsection{Position angle}

We count the number of RR~Lyrae stars and Cepheids in boxes of 0.2~kpc $\times$ 0.2~kpc in a Cartesian X,Y coordinate system projected on the equatorial plane. The coordinates were transformed using the relations from \citet{Subramanian12}, following \citet{Marel01a} and \citet{Weinberg01}. Moving along the X-axis the position of the fields with the highest number density are fitted with a first order polynomial, using the center by \citetalias{Gonidakis09} as origin.

The position angles of the young and old population of the SMC are similar. While we obtain $\Theta = 66^\circ \pm 15^\circ$ for the Cepheids, a position angle of $\Theta = 83^\circ \pm 21^\circ$ is found for the old population represented by the RR~Lyrae stars. 

Since the RR~Lyrae stars are distributed very homogeneously, the position angle does not change if we consider only stars located inside the innermost $3^\circ$ around the center of \citetalias{Gonidakis09}. For the Cepheids we find a shift of the position angle to $\Theta = 87^\circ \pm 12^\circ$ if taking only the innermost $3^\circ$ of the SMC into account. 

The position angle of the Cepheids agrees, within the uncertainties, with the literature values for Cepheids of $58^{\circ} \pm 10^{\circ}$ found by \citet{Caldwell86} and $55^{\circ} \pm 17^{\circ}$ by \citet{Laney86}. These authors used 63 and 23 Cepheids respectively, which are distributed across the central region of the SMC. \citet{Groenewegen00} found a value of $238^{\circ} \pm 7^{\circ}$, using data from OGLE~II, 2MASS and DENIS, which is in good agreement with our result as well, given a periodicity of $\pi$ of the position angle. All the different position angles are summarized in Table~\ref{inclination_table}. The value of $58.3^\circ$  for the position angle for the RR~Lyrae stars found by \citet{Subramanian12} is also similar to the value obtained in our study, but they do not quote uncertainties for their estimate.

\subsection{Inclination angle}

\begin{table*}
 \caption{Literature values of inclination ($i$) and position angle ($\Theta$)} 
 \label{inclination_table} 
\centering
\begin{tabular}{r c c c c l}
\hline\hline 
type of Stars & $\Theta$ [degrees] & $i$ [degrees] & author \\ 
\hline 
Cepheids & $58 \pm 10$ & $70 \pm 3$ & \citet{Caldwell86} \\ 
Cepheids & $55 \pm $ 17 & $ 45 \pm 7$ & \citet{Laney86} \\ 
Cepheids & $238 \pm  7$ & $68 \pm 2$ & \citet{Groenewegen00} \\ 
Red clump & 55.5 & 0.58 & \citet{Subramanian12} \\
RR~Lyrae & 58.3 & 0.50 & \citet{Subramanian12} \\
\hline 
\end{tabular}
\end{table*}

The inclination angle of the SMC is investigated by subdividing the observed fields in boxes of $0.3^{\circ} \times 0.3^{\circ}$ in $\alpha$ and $\delta$, respectively. For each field the mean distance is determined. These values are fitted by a linear function to estimate the inclination angle. 

For the RR~Lyrae stars we find a very small inclination angle of $i = 7^\circ \pm 15^\circ$. This is consistent with no inclination of the old population of the SMC whatsoever, as found by \citet{Subramanian12}. Contrary to the RR~Lyrae stars the distribution of the Cepheids is clearly inclined with respect to our line of sight. For the western parts we find mean distances for the Cepheids that are about 15~kpc farther away from us than the eastern parts of the SMC. Overall this results in an inclination angle of $i = 74^\circ \pm 9^\circ$, in very good agreement with the literature values of $68^{\circ} \pm 2^{\circ}$ from, e.g., \citet[][Table~\ref{inclination_table}]{Groenewegen00}. 


\subsection{Depth}

Several investigations have shown that the SMC has a considerable depth \citep[e.g.,][]{Mathewson88, Hatzidimitriou89, Crowl01, Subramanian12}. To determine the depth of the SMC, we use the orientation of the Cepheids and RR~Lyrae stars as seen on the sky and do not rotate the stars to align the major axes of the different populations. This approach is taken to keep the data comparable to other investigations that did not carry out such rotations either. 

\begin{figure}
\centering 
 \includegraphics[width=0.47\textwidth]{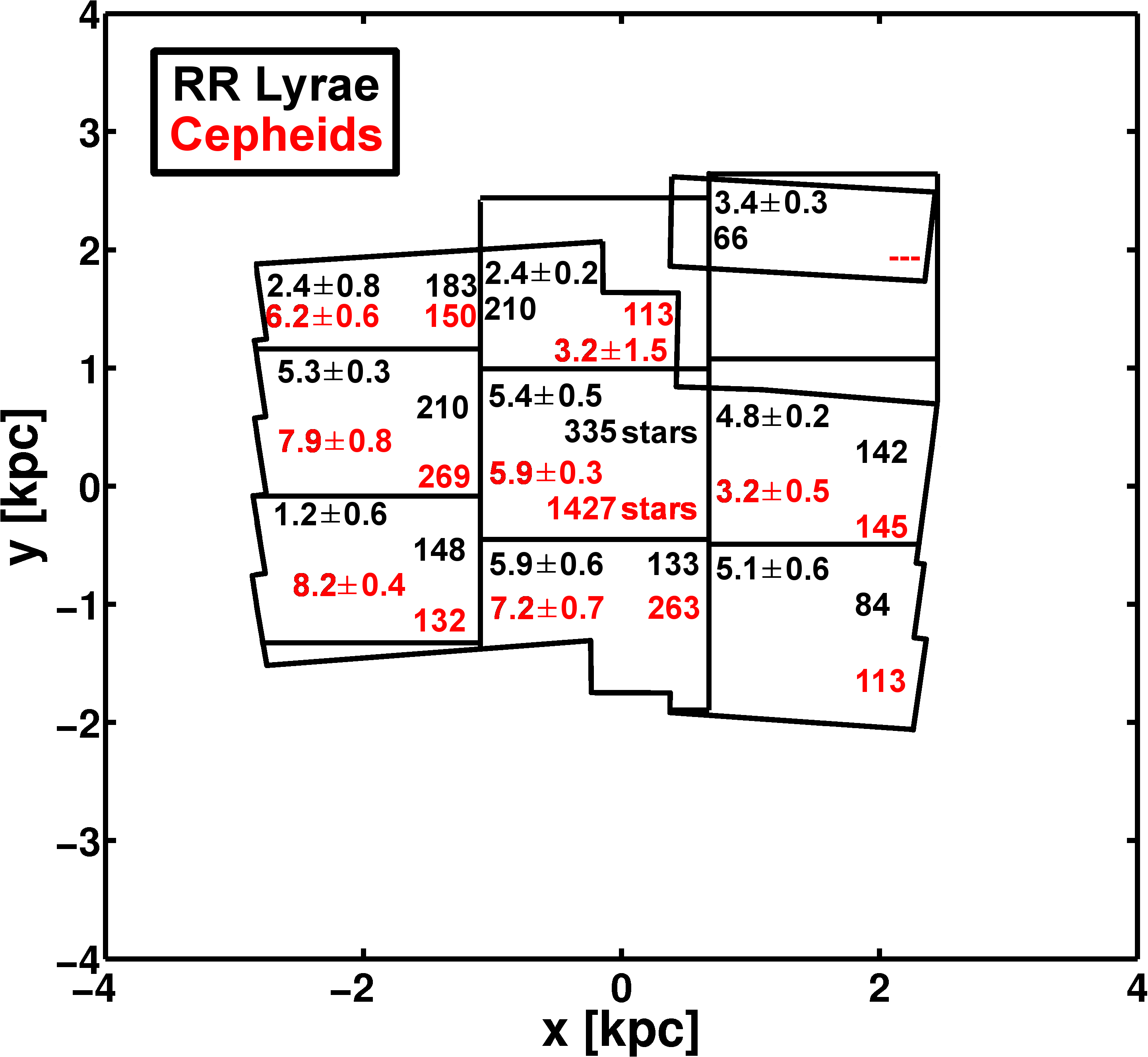}
\vspace{\floatsep}
 \includegraphics[width=0.47\textwidth]{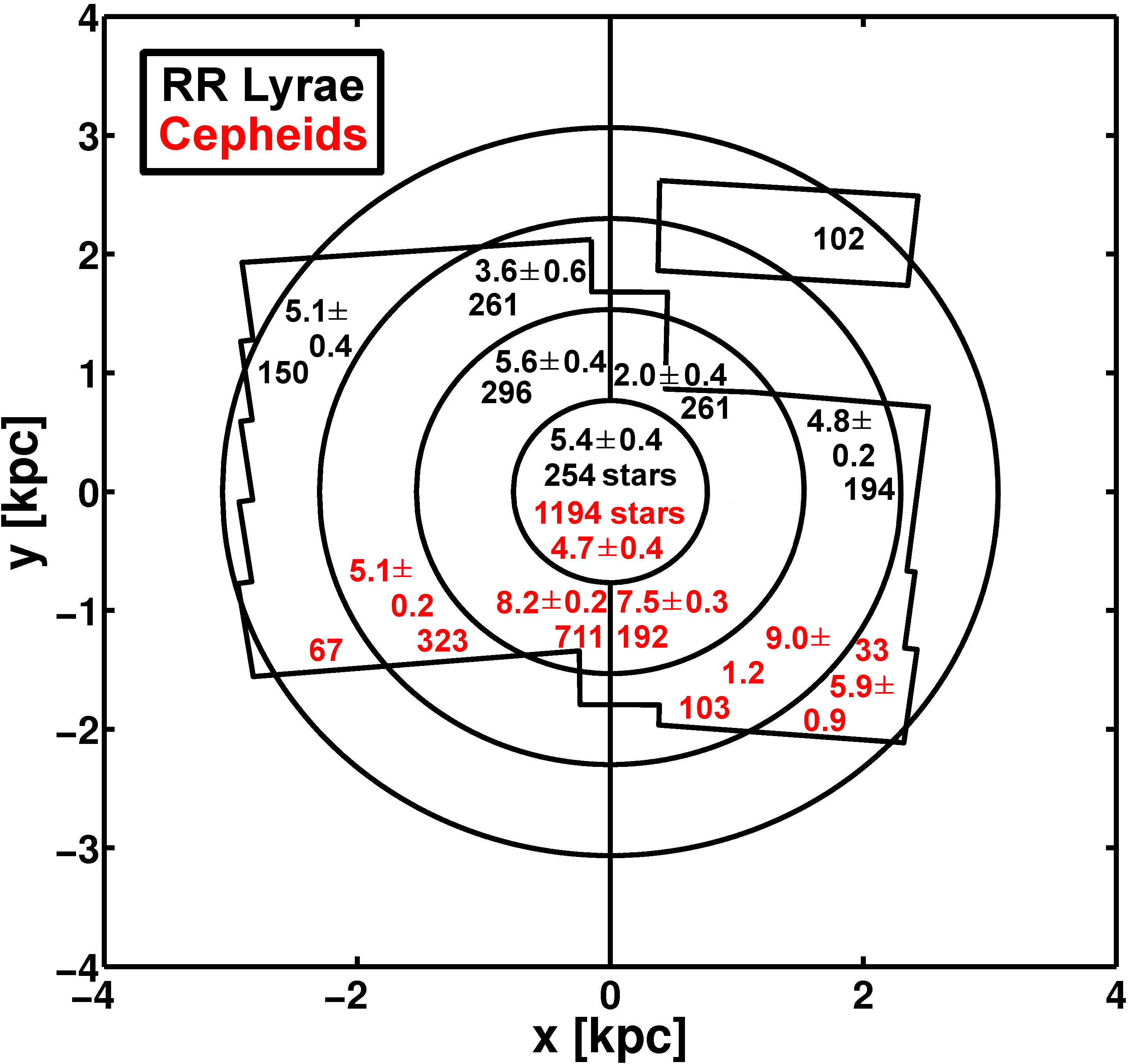}
 \caption{Distribution of depth for the RR~Lyrae stars (black numbers) and Cepheids (grey numbers, red in the online version) in the SMC. In the upper panel, the depth is calculated in the roughly rectangular fields indicated in the plot. In the lower panel semi-annular fields were used. In each field we quote the resulting depth and its uncertainty in kpc followed by the number of stars used for the calculation. A mean depth of $4.2 \pm 0.4$~kpc is found for the RR~Lyrae stars, while the Cepheids' mean depth varies around 6.0~kpc, depending on the choice of subfields.} 
 \label{scale_height_SMC} 
\end{figure}

To quantify the different depths of stars within the observed field of the SMC (compare Figure~\ref{Cep_dist_45_75}) we divide the OGLE~III field into nine rectangular fields, as well as into four rings (Figure~\ref{scale_height_SMC}). The depth of the SMC is determined by calculating a cumulative distribution of all stars present in the evaluated field. The distance values where 16\% and 84\% of all stars, respectively, have shorter individual distances than the remaining stars of the distribution, are assumed to be the lower and upper limit of the depth. These limits define the innermost 68\% of the population. The depth is determined in each of these fields individually. To compute uncertainties for the depth we vary the upper and lower limit of the depth by 5\% each and take the mean difference as the error estimate.

The measured, raw depth of the SMC is still affected by the uncertainties of the distance estimates, $\sigma_{star}$, which we subtract in quadrature to obtain the true depth. Our definition of depth is such, that it is equivalent to two standard deviations from the mean or $2\sigma_{star}$. For the RR~Lyrae stars a distance uncertainty of 3.4~kpc is found, while the Cepheid distances have an uncertainty of 3.0~kpc. Using 
\begin{equation}
depth = \sqrt{depth_{raw}^2 - (2\sigma_{star})^2}
\label{real_depth}
\end{equation}
we obtain the real depth of the SMC. For some fields the uncertainty of the distance is larger than the inferred raw depth. For these fields we do not derive a  true depth estimate (Figure~\ref{scale_height_SMC}).

The number of RR~Lyrae stars varies from field to field and decreases steadily towards the outskirts of the OGLE~III field. For the rectangular fields the depth is quite similar for all the fields chosen, as shown in the upper panel of Figure~\ref{scale_height_SMC}. The corrected depth ranges from 1.2~kpc to 5.9~kpc, with a mean depth of $4.2 \pm 0.4$~kpc. The lower panel of Figure~\ref{scale_height_SMC} shows the annular fields, divided into semi-annuli by considering only stars with either positive or negative x-values for the non-central fields. The innermost field is simply a circle around the origin. The corrected depth values reach as much as  5.6~kpc, while a mean value $4.2 \pm 0.3$~kpc is found. The northwestern fields show a slightly reduced depth, but overall no trend of differing depth within the OGLE~III field of the SMC is seen. The density of stars in the outer fields is not significantly reduced and we confirm the homogenous distribution of RR~Lyrae stars in the SMC.

For the Cepheids in the rectangular fields we find a mean corrected depth value of all fields of $5.4 \pm 1.8$~kpc. The separate field to the northwest contains only one Cepheid, and is thus excluded. The semi-annuli have a mean corrected depth value $6.2 \pm 1.8$~kpc. These mean values are close to the depth inferred for the central rectangular or circular field, where most of the Cepheids are concentrated. Evaluating the whole OGLE~III field without subdividing it leads to a (mean) depth of $7.5 \pm 0.3$~kpc.

\subsection{Scale Height}
\label{scale_height}

The depth can easily be transformed into a scale height. The scale height is the point where the density of stars has dropped by a factor of $1/e$. Therefore this quantity is half of the innermost $63\%$, instead of the $68\%$ used for the depth. Therefore the equation
\begin{equation}
 \frac{\textrm{scale~height}}{\textrm{depth}} = \frac{1 - 2(\frac{1}{2e})}{2 \times 0.68} = 0.4648.
\end{equation}
gives the transformation from the depth to the scale height. Using this transformation we obtain mean scale heights of $2.0 \pm 0.4$~kpc for the RR~Lyrae stars using the rectangular and the semi-annular fields. For the Cepheids the different field selections lead to different scale height estimates. They range from $2.5\pm 0.4$~kpc for the rectangular fields to $2.9 \pm 0.3$~kpc when evaluating the whole OGLE~III field.

Several estimates of the depth have been provided in the literature leading to very different pictures of the SMC. Using 61 Cepheids across the bar of the SMC \citet{Mathewson88} found a depth of 20~kpc, twice as much as found in our work. Using RC stars \citet{Subramanian09, Subramanian12} concluded for the OGLE~II and OGLE~III fields, respectively, that the 1-$\sigma$ depth is below 5~kpc. They used the width of the distribution of magnitudes of the RC stars to estimate the depth. This width is influenced by several different factors, such as reddening, metallicity differences, or true depth effects, possibly leading to an underestimate of the actual depth. The approach taken by \citet{Subramanian09} for the RC stars was used by \citet{Kapakos11} for a subset of about 100 RR~Lyrae stars present in the bar region of the OGLE~III survey and for the whole OGLE~III RR~Lyrae stars dataset by \citet{Subramanian12} assuming reddening values from RC stars. Both investigations found a 1-$\sigma$ width of the distribution of stars of $\sim 4$~kpc, in good agreement with our results for the RR~Lyrae stars. However, \citet{Subramanian12} conclude that the depth could be as much as 14~kpc taking 3.5-$\sigma$ of the distribution into account. Too few stars and too small a field were investigated in \citet{Kapakos11} to determine spatial differences for the line-of-sight depth. Using the complete OGLE~III dataset of the RR~Lyrae stars, \citet{Subramanian12} investigated the depth of 70 very small fields and found that the northern and eastern parts may have a slightly decreased depth. This is in good agreement with our results. Furthermore, the estimates of the depth relying on cluster distances agree very well with our calculations for the young population traced by Cepheids. \citet{Crowl01} found a depth of 6~kpc to 12~kpc from ground-based imaging, while the six intermediate-age clusters studied by \citet{Glatt08b} with deep \textit{HST} imaging lead to a depth of $\sim 10$~kpc, excluding the star cluster NGC~419, which seems to be 6~kpc closer than the closest of the other six clusters. 

%

\section{Summary and Conclusions}
\label{Conclusions}

We investigate the three-dimensional structure of the young and old population of the SMC by calculating individual distances to 2522 Cepheids and 1494 RR~Lyrae stars. The absolute magnitudes of the RR~Lyrae stars are calculated from the photometric metallicity estimates in \citet{Haschke12_MDF}. These are based on the Fourier-decomposed lightcurves of the RR~Lyrae~\textit{ab} stars observed by the OGLE~III survey. The period data to compute absolute magnitudes for the Cepheids are taken from the dataset of the OGLE~III survey as well.

We use two different approaches to correct for the reddening that the investigated stars are experiencing. First the differences between the observed and theoretical mean color of red clump stars are used to estimate an area-averaged reddening value. For the other technique individual reddening estimates of each Cepheid and RR~Lyrae star are calculated. With this method we are able to correct the actual line-of-sight reddening at the very position of the star. The reddening maps of both techniques are shown in \citet{Haschke11_reddening}. These result in two sets of self-consistent three-dimensional maps of the SMC for the young Population~I (Cepheids) and the old Population~II (RR~Lyrae) stars.

\begin{table*}
\caption{Results for the distances and structural parameters of the SMC.} 
 \label{Summary_table} 
\centering       
\begin{tabular}{r c c c c}  
\hline\hline     
 & RR~Lyrae & Cepheids \\ 
\hline      
distance modulus using area averaged reddening & $19.13 \pm 0.13$ & $19.17 \pm 0.12$ \\ 
distance modulus using individual reddening & $18.94 \pm 0.11$ & $19.00 \pm 0.10$ \\ 
inclination [degree] & $7 \pm 15$ & $74 \pm 9$ \\ 
position angle [degree] & $83 \pm 21$ & $66 \pm 15$ \\ 
scale height [kpc] & $2.0 \pm 0.4$ & $2.7 \pm 0.3$ \\ 
\hline         
\end{tabular}
\end{table*}
Using individual reddening estimates we calculate a median distance modulus of $(m-M)_{0 \mathrm{RRL/median}} = 18.94 \pm 0.11$ for the RR~Lyrae stars and $(m-M)_{0 \mathrm{Cep/median}} = 19.00 \pm 0.10$ for the Cepheids. The results of the median distance moduli are in very good agreement with each other and with the distances obtained in the literature (Table~\ref{distance_table} and Table~\ref{Summary_table}). By applying reddening values obtained with the area-averaged reddenings we find distances that are a bit larger, but still in good agreement with many earlier distance estimates (Table~\ref{distance_table}). We explain the larger distance values of the area-averaged extinction technique with unresolved differential reddening effects \citep[e.g.,][]{Barmby11}. We do not find any evidence for a \textquotedblleft long\textquotedblright \ and \textquotedblleft short\textquotedblright \ distance scale problem when comparing our RR~Lyrae and Cepheid distances to the SMC. A similar result was found for the LMC \citep{Haschke12_LMC}. 

The RR Lyrae stars show a fairly homogeneous distribution across the OGLE field. Their density increases gradually towards the center of the field, while the highest-density regions are located in a double-peaked, semi-annular structure around the center \citep[see also][]{Soszynski10b, Subramanian12}. Overall the RR Lyrae stars show a roughly spheroidal or ellipsoidal distribution. This spheroidal distribution becomes more pronounced and easier recognizable when taking the much more numerous intermediate-age populations into account \citep[not studied in our paper, but see, e.g.,][]{Zaritsky00, Cioni00a, Gonidakis09, Subramanian12}. The RR Lyrae stars do not reveal any obvious correlations with younger irregular features such as the SMC bar and there is no bar visible in the distribution of RR~Lyrae stars. While our analysis is only mildly suggestive of the eastern part of the old population's distribution being closer to us, this trend is confirmed more clearly when larger areas than the OGLE~III field are taken into account, such as in the sparsely sampled survey of the outer SMC regions by \citet{Nidever11}, or more numerous intermediate-age populations such as RC stars \citep[e.g.,][]{Hatzidimitriou89, Gardiner91}. Moreover, the kinematics of intermediate-age and old red giants across the central parts of the SMC suggest that they are part of an unperturbed pressure-supported spheroid \citep{Harris06}. 

In contrast, the distribution of the Cepheids is closely correlated with the regions of recent star-formation activity along the bar of the SMC. The wing of the SMC, another region of ongoing star formation, lies outside of the OGLE~III field. We emphasize that the Cepheids are tracers of a slightly older population (some 30 -- 300 Myr) than the one responsible for the prominent H~{\sc ii} regions along the SMC bar \citep[see, e.g., the images in][]{Bolatto11}. Using Cepheids as tracers, we find the eastern part of the SMC field with distances $< 55$~kpc to be closest to us, in good qualitative agreement with the results from older tracers. Cepheids at distances between $\sim 55$ and $\sim 60$ kpc are still found mainly in the eastern part of the OGLE~III field, where they coincide with the north-eastern part of the bar around the luminous H~{\sc ii} region N66. At distances starting at $\sim 62$ kpc most of the Cepheids are concentrated around the SMC center derived by \citetalias{Gonidakis09} from older K and M giants in 2MASS. This region overlaps in projection with N17 and its neighboring H~{\sc ii} regions in the lower (southwestern) region of the bar. At distances in the range of $\sim 65$ to $\sim 68$~kpc we still find the highest concentration of Cepheids near \citetalias{Gonidakis09}'s center, with a less prominent, scattered tail extending further east and a sparse scattering of stars to the west. Almost no Cepheids are observed in the northwestern part of the bar at these farther distances. If we assume that the Cepheids are physically associated with the bar, this indicates that the bar is tilted from the northwest (closest part) to the southeast (farthest part, elongated along the line of sight). This is visualized in the included mpeg movie.

In their analysis of the global star formation history of the SMC \citet{Harris04} inferred that the SMC was comparatively quiescent at intermediate ages (about 8.4 to 3 Gyr ago), while the star formation activity increased at more recent times. They find peaks at 2.5 and 0.4 Gyr, which they attribute to close encounters of the SMC with the Milky Way (in agreement with other studies), and a most recent peak at 60 Myr. The latter two maxima roughly bracket the ages of Cepheids. \citeauthor{Harris04}'s distribution of star formation activity at 400 Myr and 250 Myr in their Figure 6 resembles the distribution of Cepheids found in our study. Whether indeed tidally triggered star formation created the Cepheids is unclear since recent high-precision proper motion measurements for the Magellanic Clouds raised new questions regarding their short- and long-term orbital history \citep[e.g.,][]{Kallivayalil06b, Besla07}. We note that the two-dimensional distribution of star clusters in the age range of the Cepheids coincides well with the general locus of the Cepheids, although the star clusters are more strongly confined to the bar \citep[see][their 
Figure 8]{Glatt10}. 

The position angle of the two populations are similar. For the Cepheids we obtain $\Theta_{\mathrm{Cep}} = 66^\circ \pm 15^\circ$, while $\Theta_{\mathrm{RRL}} = 83^\circ \pm 21^\circ$ is found for the RR~Lyrae stars. Unlike the position angle the inclination angle changes significantly between the old and young population. We find an inclination angle of $i_{\mathrm{RRL}} = 7^\circ \pm 15^\circ$ for the RR~Lyrae stars, which is consistent with zero inclination. For the Cepheids an inclination of $i_{\mathrm{Cep}} = 74^\circ \pm 9^\circ$ is obtained such that the northeast is much closer to us than the southwest as mentioned earlier. The closer part is roughly pointing towards the LMC. We visualize the three-dimensional distribution of the Cepheids and RR~Lyrae stars in the mpeg movie in Figure~\ref{RRL_Cep_dist_colorext}. Overall, the comparison of the structural parameters from Cepheids and RR~Lyrae stars found in the literature to the results found in this study yields good agreement (Table~\ref{inclination_table} and Table~\ref{Summary_table}). 

The OGLE~III field is too small to deduce the actual shape of the SMC, which remains under debate. Only large scale surveys of the whole SMC including the outskirts will solve this issue.

Nonetheless, we can infer depth estimates from our investigation. The depth of the SMC has been under extensive discussion for the last decades. \citet{Mathewson88} claimed the SMC to be very extended with a depth of 20~kpc, while \citet{Subramanian09, Subramanian12} and \citet{Kapakos11} found a 1-$\sigma$ depth of less than 5~kpc. We find different depths for the old population (RR~Lyrae stars) and for the young population (Cepheids). For the RR~Lyrae stars we find a 1-$\sigma$ depth of $4.2 \pm 0.4$ (or a scale height of $2.0 \pm 0.4$~kpc), while the depth for the Cepheids is measured, depending on the field selection, to be between $5.4 \pm 1.8$~kpc and $6.2 \pm 0.3$~kpc (or a scale height of $2.7 \pm 0.3$~kpc). 

Usually the scale height for the young population is expected to be smaller than for old populations. 

However, in the SMC the young population clearly has a very different distribution than the old population, showing an asymmetric and highly inclined distribution Although there is considerable uncertainty regarding the long-term orbits of the SMC, LMC, and Milky Way it seems quite likely that the recent, increased star formation leading to the Cepheids was triggered by a close encounter between these galaxies. This encounter may have shifted and compressed some of the SMC's gas through tidal and ram pressure effects, possibly even creating some of the features that we observe now as the bar and the wing.

%
%
\acknowledgments
We thank our anonymous referee for his or her helpful comments. We are thankful to the OGLE collaboration for making their data publicly available. R. Haschke is obliged to S. Schmeja for giving helpful comments on the calculation of the $\mathcal{Q}$-parameter. The comments to improve the manuscript and the proof reading by K. Glatt and K. Jordi are very much appreciated. This work was supported by Sonderforschungsbereich SFB~881 ``The Milky Way System'' (subproject A2) of the German Research Foundation (DFG).

\bibliography{Bibliography.bib}
\bibliographystyle{apj}

\end{document}